\documentstyle[12pt]{article}
\begin{document}
\vskip 2.5cm

{\renewcommand{\thefootnote}{\fnsymbol{footnote}}
\centerline{\large \bf  A Simple Ansatz to Describe Thermodynamic Quantities}
\centerline{\large \bf  of Peptides and Proteins at Low Temperatures}
\vskip 3.0cm

\centerline{Ulrich H.E.~Hansmann$^{1}$
\footnote{\ \ e-mail: hansmann@ims.ac.jp}
 and Philippe de Forcrand$^{2}$
\footnote{\ \ e-mail: forcrand@scsc.ethz.ch}}
\vskip 1.5cm
\centerline {$^{1}$ {\it Department of Theoretical Studies,
Institute for Molecular Science}}
\centerline {{\it Okazaki, Aichi 444, Japan}}
\centerline{$^{2}$ {\it Swiss Center for Scientific Computing (SCSC)}}
\centerline{ \it Eidgen{\"o}ssische Technische Hochschule (ETH) Z{\"u}rich,
8092 Z{\"u}rich, Switzerland}

\medbreak
\vskip 3.5cm
 
\centerline{\bf ABSTRACT}
\vskip 0.3cm
We describe a simple ansatz to approximate the low temperature behavior
of proteins and peptides by a mean-field-like model which is
analytically solvable. For a small peptide some thermodynamic quantities
are calculated and compared with numerical results of an all-atoms simulation.
Our approach can be used to determine the weights for a multicanonical
simulation of the molecule under consideration.
\vskip 3.5cm

\noindent
{\it Key words:} Mean-field-like model for proteins, Multicanonical weights.
 
\vfill
\newpage}
\baselineskip=0.8cm
Prediction of the three-dimensional structure of peptides and proteins 
solely from their amino-acid
sequence remains one of the longstanding unsolved problems in  bioscience.
  It is widely accepted that 
the three-dimensional shape at temperature $T$ 
 corresponds to the global minimum in {\it free} energy of the molecule.
Therefore, given a sufficiently accurate description of the intramolecular 
forces, 
 it should be  in principle possible to predict such conformations through
a numerical simulation. However, the complex form of the interactions
containing both repulsive and attractive terms leads to a very rough 
 energy landscape with  
 a huge number of local minima separated by high energy barriers.
These barriers slow down the ergodic exploration of conformation space at low 
temperatures.
In the canonical ensemble at temperature $T$, the  probability to cross an 
energy barrier of height $\Delta E$ is proportional to $e^{-\Delta E/k_B T}$.
Hence, at low temperatures, canonical molecular dynamics and Monte Carlo, 
using local updates, will get trapped in  one of these local minima.
Within limited simulation time,
only small parts of phase space are sampled and physical quantities
cannot be calculated accurately. 

However, with the development of the multicanonical approach \cite{MU},
simulated tempering \cite{LMP} and other {\it generalized-ensemble} 
techniques, an efficient sampling of low energy configurations and 
an accurate calculation of thermodynamic quantities at low temperatures became
feasible, at least for small peptides. The first application of one  such
technique  to the protein folding problem can be found in Ref.~\cite{HO}. 
Later applications include the study of the coil-globular
transitions of a model protein \cite{HSp} and the helix-coil transitions of
homo-oligomers of nonpolar amino acids \cite{HO95a}. A numerical comparison
of several generalized-ensemble algorithms can be found in Ref.~\cite{HO96b}.
Nevertheless, calculation of thermodynamic quantities by numerical
simulations remains a time consuming process. 
The evaluation of energies which is necessary for each Monte Carlo or molecular 
dynamics step is itself CPU time intensive.
Hence, it is desirable to find an approximate description of the
 thermodynamic properties of complicated molecules like peptides
 by using a much simpler effective model. The form of the effective 
 model should retain the essential physical characteristics
of the original system (number and nature of the degrees of freedom)
and its parameters be adjusted to best reproduce the thermodynamic 
properties of the original, especially at low temperature.
 The results from generalized 
ensemble simulations now allow the development and testing
of such an approach.

Fixing bond lengths and bond angles (which within our range of temperatures
can be considered as rigid),
peptides and proteins can be described by a set of torsion angles. This is
the motivation behind our choice to describe our effective model by 
$n_F$ {\em independent} angles $\theta_i, i=1,...,n$, where 
$\theta_i \in [-\pi,\pi[$. 
We assume that the potential
energy of our system can be approximated by
\begin{equation}
E \approx \sum_i^{n_F} V (\theta)~,
\label{eq4}
\end{equation}
where  the ``mean'' potential $V(\theta)$ is the same for each angle.
The most general form for the periodic potential $V$ is
\begin{equation}\label{full}
V(\theta) = \sum_{k=0}^{\infty} a_k \cos k \theta  + b_k \sin k \theta
\end{equation}

Without loss of generality, we can impose that the global minimum 
of $V$ occurs, for instance, at $\theta=0$. Then the low-temperature 
behavior will be determined by the harmonic approximation
\begin{equation}
V_{harmonic}(\theta) = c_0 + c_1 (1 - \cos \theta)~.
\end{equation}
To maintain the simplicity of our effective model, we truncate 
the expansion in Eq.~(\ref{full}) to 
\begin{equation}
V(\theta) = c_0 + c_1 (1 - \cos \theta) + c_2 (1 - \cos \theta)^2
\label{eq5}
\end{equation}
with anharmonic term $c_2$. The 3 parameters $c_0, c_1,$ and $c_2$ 
will be fitted to best match features of the original model.
It is obvious that 
\begin{equation}
c_0 = \frac{E_{GS}}{n_F}
\label{eq6}
\end{equation}
where $E_{GS}$ is the ground state energy of our molecule. 

The partition function of our effective model is simply
\begin{equation}
Z = z^{n_F} \quad 
{\rm with} \quad z = \int_{-\pi}^{+\pi} d\theta e^{-\beta V(\theta)}
\label{Eqpar}
\end{equation}
Using the definition of the modified Bessel function 
$I_n(y) = \frac{1}{\pi} \int_0^{\pi} dx~\cos nx~e^{y~\cos x}$,
and the identity $e^{y~\cos x} = I_0(y) + 2 \sum_{k=1}^{\infty} 
I_k(y)~\cos kx$,
the partition function per angle $z$  can be expressed as
\begin{equation}
z = 2 \pi e^{\displaystyle -\beta (c_0 + c_1 + \frac{3}{2} c_2)}
\left( I_0(-\beta c_2/2) I_0\left( \beta (c_1 + 2 c_2)\right) 
+ 2 \sum_{k=1}^{\infty} I_k(-\beta c_2/2) I_{2k}\left( \beta (c_1 + 2 c_2)
\right) \right)
\label{Pf}
\end{equation}

The average energy $<E>$  can be obtained through 
$<E> = - d(\log Z) / d\beta$, or directly as
\begin{equation}
<E> =\left[ (c_0 + c_1 + \frac{3}{2} c_2) - (c_1 + 2 c_2) <cos x> 
                                    + \frac{c_2}{2} <cos 2x>\right] n
\label{ave}
\end{equation}
with
\begin{eqnarray}
<\cos mx> & = & f(m) / f(0) \\
f(m) & \equiv & I_0(-\beta c_2/2) I_m\left( \beta (c_1 + 2 c_2)\right)\\
& & 
+ \sum_{k=1}^{\infty} I_k(-\beta c_2/2) 
\left[ I_{2k+m}\left( \beta (c_1 + 2 c_2)\right) 
                      + I_{2k-m}\left( \beta (c_1 + 2 c_2)\right) \right]
\end{eqnarray}
and similarly for the specific heat  which we define as
\begin{equation}
C(T) = \frac{1}{k_B n_A} \frac{{\rm d}}{{\rm d}T} <E>
\label{sph}
\end{equation}
with Boltzmann constant $k_B$ and $n_A$ the number of residues.

We expect 
Eq.~\ref{eq4}, with the mean potential given by Eq.~\ref{eq5}, to be
a good approximation of our system for low temperatures. In 
principle the approximation can be systematically improved  for higher temperatures 
by increasing  the number  of terms in the expansion of Eq.~\ref{full}.
However, this will lead to more complicated equations than for instance
Eqs.~\ref{Pf} and \ref{ave}.
While periodicity of the variables $\theta$ is not required for a
low-temperature approximation, this similarity with the original system
improves the quality of the approximation at higher temperatures (e.g.
the specific heat remains bounded).

Our benchmark is  one of the simplest peptides, 
Met-enkephalin, which has become  an often used
model to examine new algorithms in the protein folding problem. 
Met-enkephalin has the amino acid sequence Tyr-Gly-Gly-Phe-Met. For
comparison and to fit the free parameters of our ansatz we use the 
results published in Ref.~\cite{HE96c}. They were derived from a
multicanonical simulation of 200,000 sweeps using a 
 potential energy function $E_{tot}$  given 
by the sum of 
electrostatic term $E_C$, Lennard-Jones term $E_{LJ}$, and
hydrogen-bond term $E_{hb}$ for all pairs of atoms in the peptide
together with the torsion term $E_{tors}$ for all torsion angles.
The parameters for the energy function were adopted from
ECEPP/2 \cite{EC3}.  By fixing the peptide bond angles $\omega$ to 
$180^{\circ}$,  19 torsion angles were left as degrees of freedom
(i.e.~$n_F=19$).

In our ansatz using Eq.~\ref{Pf} to approximate our molecule 
we have  three free parameters which
 have to be fitted against numerical results.
Since our approximation is only valid for low temperatures, we should
perform the  fit against results at  the lowest reliable temperatures
available. We expect canonical simulations of peptides and proteins to be
computationally feasible down to the folding temperature $T_F$.
A fit of our free parameters should therefore be done against 
thermodynamic quantities measured in the vicinity of this temperature. 
By using results from simulated annealing
runs it may be possible to include results of even lower temperatures.
In our case, however, we  restricted ourselves to the temperature range
$200~K$ to $240~K$, since the folding temperature was found to be 
$T_F\approx 220~K$ for Met-enkephalin \cite{HO97b}.
For our fit in this temperature range we use results 
for the average energy $<E>$ and specific heat $C(T)$ as obtained from 
a multicanonical simulation and published in Ref.~\cite{HE96c}.  
We find $c_0= -0.55$ kcal/mol, $c_1 = 4.77$ kcal/mol and 
$c_2 = -2.05$ kcal/mol. 
These parameters were obtained in the following way. Since one shortcoming of
our ansatz is to underestimate the peak in the specific heat, we first fixed
the ratio $c_2 / c_1$ to the value $-0.43$ which gives the highest specific 
heat peak. Then we adjusted $c_1$ to reproduce the specific heat in the chosen
temperature range $200~K$ to $240~K$.
Finally the remaining
parameter $c_0$ was fixed by requiring that our fit reproduce the 
values for the average energy $<E>$ in the chosen temperature range.  For the
case of a pure harmonic approximation, i.e. fixing $c_2 = 0$, we were not
able to obtain a fit which would reproduce $<E>$ and $C(T)$ in the temperature
range $200~K$ to $240~K$.  The closest low-temperature match was obtained 
for a choice of 
parameters $c_0=-0.55$ kcal/mol and $c_1 = 1.19$ kcal/mol.
We remark that  $n_F c_0$ is an estimate for the ground state energy. Hence,
the value of $n_F c_0 \approx -10.5$ kcal/mol should
be compared with the known true value of $E_{GS} = -10.72$ kcal/mol.

Using the above values for $c_i$ we were now able to calculate 
estimates for the average energy $<E>$ and specific heat $C(T)$ by our ansatz 
and compare the obtained results with those of a multicanonical 
simulation. In Fig.~1 we display $<E>$ and in 
Fig.~2 the specific heat $C(T)$ as obtained by our ansatz (with and without
anharmonicity constant $c_2$) and by a multicanonical simulation. As one can 
clearly see from our data, it is necessary to include anharmonic
terms in our ansatz. A pure harmonic approximation  produces reasonable 
values for the average energy $<E>$  only for temperatures
below $150~K$ and for the specific heat below $T\sim90~K$.
Including an anharmonicity constant $c_2$  dramatically improves
the ability of our ansatz to predict average energies. Values for this
quantity can now be reproduced up to temperatures  $\approx 300~K$,  
i.e. to temperatures  where usual canonical Monte Carlo simulations
are feasible. However,  our approach is less successful for the
specific heat, where the data from the multicanonical simulation can
only be reproduced up to temperatures  $\approx 220~K$, the folding temperature
$T_F$.  
Better agreement could be obtained with the addition of further anharmonic 
terms.

We have demonstrated that for small peptides the low temperature behavior of 
energy and related quantities can be described by our simple ansatz. In 
addition, our ansatz can also be used to obtain estimators for the weights 
in generalized-ensemble simulations. Unlike for the canonical ensemble
 these weights are not {\it a priori} known,
and a great fraction of the computing effort (about 1/2) goes into their
determination prior to the simulation itself. 
As an example
take the multicanonical algorithm. Here, the weights are defined by
\begin{equation}
w_{mu}(E) \propto n^{-1}(E) =  e^{-S(E)}~,
\end{equation}
where $n(E)$ is the spectral density and $S(E)$ the microcanonical entropy.
Obviously, knowledge of the weights in this ensemble is equivalent to
that of the spectral density, i.e. to solving the system. Hence, the need for 
estimators. Usually these estimators are calculated by an iterative procedure,
first described in Ref.~\cite{MU}. However, convergence of this method 
depends strongly on the model under consideration, and determining the
weights can be a time consuming and difficult process. Several attempts
have been made to speed up their calculation, see for instance 
Refs.~\cite{berg, bruce,vasq}, but there is still need for further
improvement.
Here, we propose to take as estimators for the multicanonical weights 
\begin{equation}
w_{mu} (E) \propto \tilde{n}^{-1} (E) = e^{- \tilde{S}(E)}~,
\label{eq3}
\end{equation}
where $\tilde{n}(E)~(\tilde{S}(E))$ is the  spectral density (microcanonical 
entropy) of our effective model. 
Note that once multicanonical weights are
known, it is easy to calculate the weights for other 
generalized ensemble techniques  as was shown in Ref.~\cite{HO96c}. 
In principle, $\tilde{n}(E)$ can be calculated
directly  from the partition function  Eq.~\ref{Eqpar}. However,
in practice it is easier to calculate the canonical entropy
\begin{equation}
\tilde{S}(T) = \frac{<E>(t)}{k_B T} + \log Z(T)~,
\label{entropy}
\end{equation}
and approximate $\tilde{S}(E) \approx \tilde{S}(T[<E>])$. 
The function $<E>(T)$ can be 
calculated by our ansatz, which allows us in turn to obtain $T(<E>)$. 
This approximation of the microcanonical entropy by the canonical one
becomes exact for $n_F \longrightarrow \infty$. 

With the above ansatz we were able to calculate new multicanonical weights
for Met-enkephalin, for energies between $-10.7$ and $0$ kcal/mol. The 
upper limit comes from the fact that our approach gives a good approximation 
of the average energy only for $T \le 300~K$,  where 
 $<E>(T=300~K) \approx 0 $.
For higher energies we set $w(E) = c \cdot e^{-E/(300 \cdot k_B)}$ where the 
constant $c$ is chosen such that the weights are a continuous function of the
energy at $E=0$.  In this way we also allow sampling of high
energy states. However, it is clear that with so defined weights 
use of the re-weighting techniques is essentially restricted to 
temperatures $T \le 300$ K. Re-weighting to higher temperatures
would lead to expectation values hampered by  increasing errorbars  since
the sampling of configuration becomes poorer  and poorer with increasing 
energies for $E > 0$.   

We used for the calculation of multicanonical weights our
full $3$-parameter ansatz. 
  since the harmonic approximation fails already 
  at much lower temperatures (and hence energies). 
In Fig.~3 we show the 
distribution in energy obtained from a simulation with 1,000,000 sweeps 
using these weights.  The distribution  covers the chosen energy range and 
is essentially flat in this range . The histogram entries for the 
energies differ by no more than a factor 2. This near-constancy demonstrates
that the entropy Eq.~\ref{entropy} of our ansatz is a very good approximation of
the exact microcanonical entropy, yielding
good estimators for the
multicanonical weights. The quality of these estimators and the range of
 energies could easily be enhanced by  further iterations with the method
described first in Ref.~\cite{MU}. However, since our simulation
covers a broad range of energies, we can use standard reweighting techniques
\cite{FS} and obtain already with the given
weights   accurate  thermodynamic quantities  over a  range
of temperatures not accessible by canonical simulations. As an example
we show in Fig.~4 the average end-to-end distance $<D>(T)$ (measured from
N of Tyr 1 to O of Met 5) as a function of temperature as obtained from
the above  multicanonical simulation with the new weights. This quantity is
a simple measure for the compactness of configurations.  For comparison we
also show the results from a multicanonical simulation  with the same number 
of sweeps but with the more constant weights used in Ref.~\cite{HE96c}. 
As one can clearly see the two sets of weights lead to  values of the 
 end-to-end distance  compatible with each other. 
Note that in the case of  the new weights the errorbars 
increase rapidly with temperature for $T > 300$ K, reflecting the fact that for
$E \ge 0$  configurations are no longer sampled according to a
uniform distribution but to a Boltzmann distribution at $T=300$ K.
On the other hand, the old set of weights
yields a flat distribution for energies up to $\approx 20$ kcal/mol (which
corresponds to $T=1000K$) and therefore the size of errorbars does not increase
with temperature in the shown range of temperatures. 

For a study of the protein folding problem one would prefer proteins which 
are at least an order of magnitude larger than Met-enkephalin, the peptide
we used in this work.  
It is difficult to predict how useful our ansatz is going to be in this case.
On the one hand one expects a statistical, mean-field description to be even 
more accurate on a larger system. On the other hand, a larger protein will have
a more pronounced folding transition, becoming first-order-like
(for a discussion on
the nature of the folding transition see, for instance, Ref.~\cite{Woln}),
and this collective phenomenon will be impossible to capture with our
mean-field approach. Therefore we would expect our ansatz to work better
below the folding temperature $T_F$, more poorly at or above it.
Its efficiency for the determination of multicanonical weights remains
to be determined.

To summarize our results, we have introduced a simple
mean-field-like model to describe the low temperature behavior of
peptides and proteins, which can be systematically improved by including
higher Fourier components in the mean-field potential. 
For a small peptide we calculated some
thermodynamic quantities and compared the results with the ones from
an all-atoms simulation. In addition, our ansatz allows  for peptides
a simple
and efficient calculation of multicanonical weights alleviating in
this way the main drawback of these techniques. 

\vspace{0.5cm}
\noindent
{\bf Acknowledgements}: \\
Our simulations were performed on the computers at SCSC of the ETH
Z{\"u}rich, Switzerland, and 
at the Institute for Molecular Science (IMS), Okazaki,
Japan.\\


\newpage
{\Large Figures:}
\begin{enumerate}
\item Average energy $<E>$ as a function of temperature.  Shown are the
      results obtained from our ansatz, for both $c_2=0$ (dot dashed line)
       and $c_2 \neq 0$ (unbroken line). 
      For comparison we also show the results obtained from a 
      multicanonical run of 200,000 sweeps (marked by $+$).
\item Specific heat $C(T)$ as a function of temperature.  Shown are the
      results obtained from our ansatz, for both $c_2=0$ (dot dashed line)
      and $c_2 \neq 0$ (unbroken line). For
      comparison we also show the results obtained from a multicanonical run
      of 200,000 sweeps (marked by $+$).
\item Multicanonical distribution of energy, 
      $P(E) \propto e^{-S(E) + {\tilde S}(E)}$, obtained from a multicanonical
      simulation of 1,000,000 sweeps with weights calculated from our ansatz.
\item Average end-to-end-distance $<D>_T$ as a function of temperature.
      The data marked by $\diamond$ are from a multicanonical simulations 
      of 1,000,000 sweeps 
      with weights obtained by our new ansatz. For comparison we also show 
      the results (marked by $+$) from a multicanonical simulation with 
      same number of MC 
      sweeps and the weights of Ref.~\cite{HE96c} which were 
      obtained by the older iterative procedure. 
\end{enumerate}
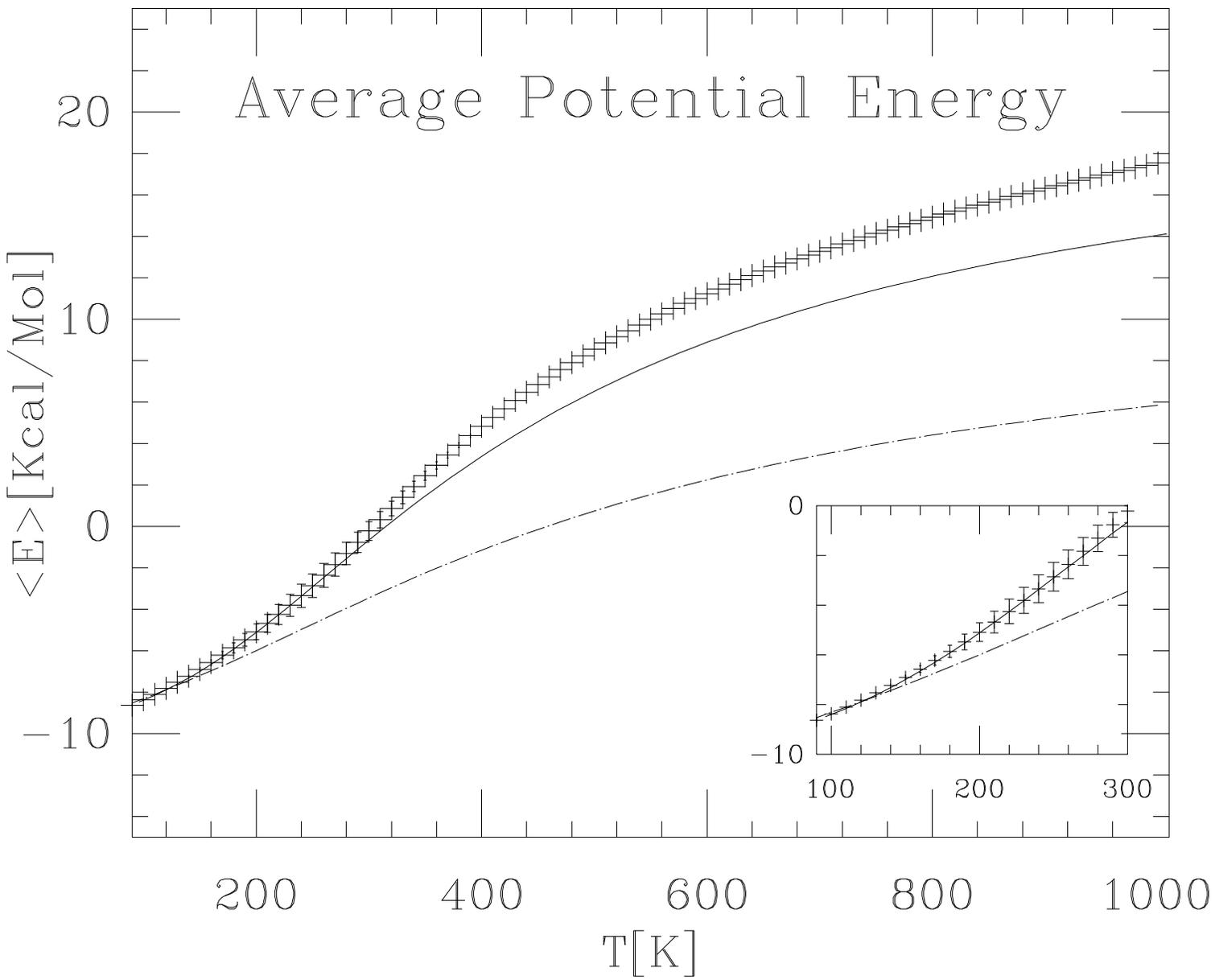
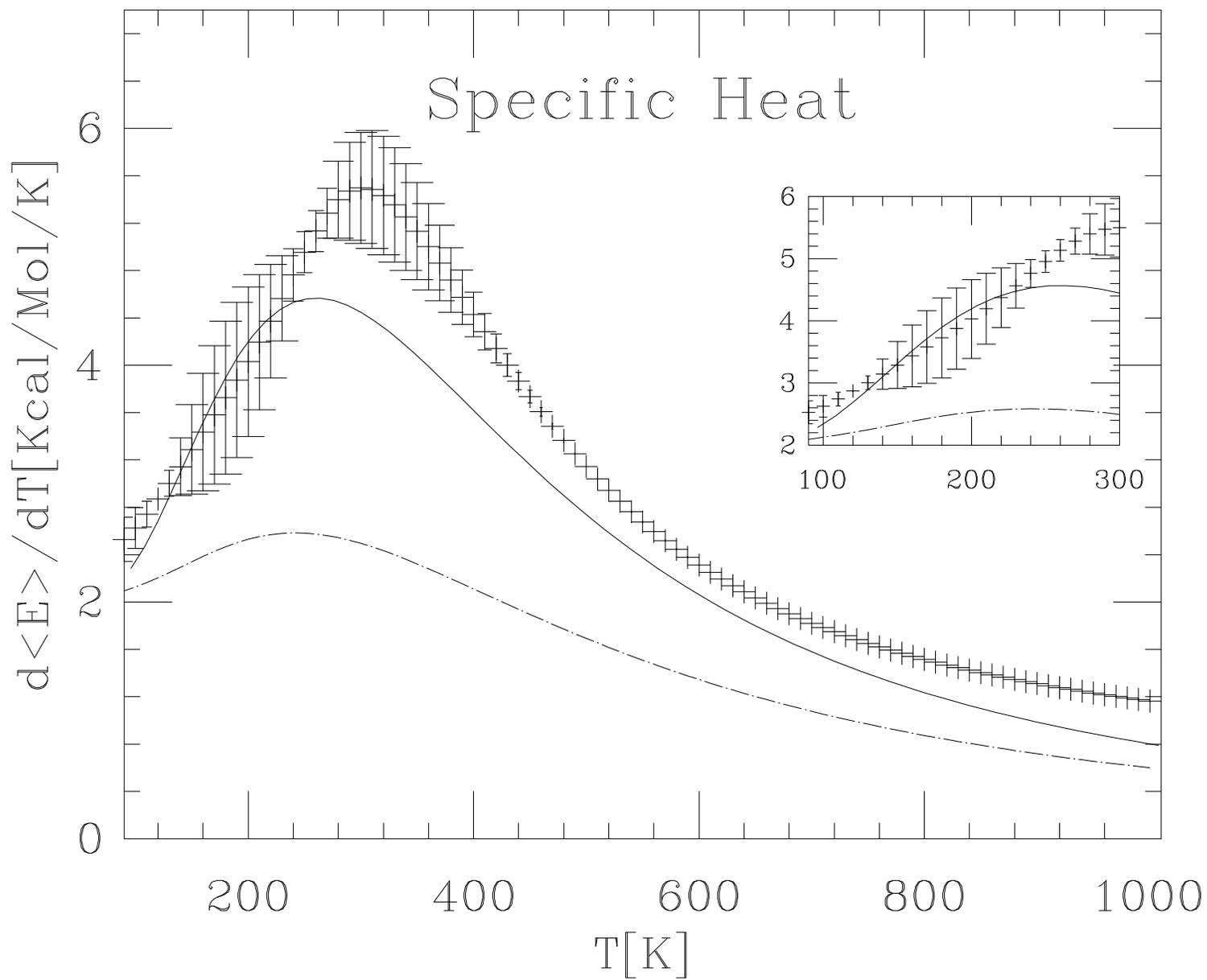
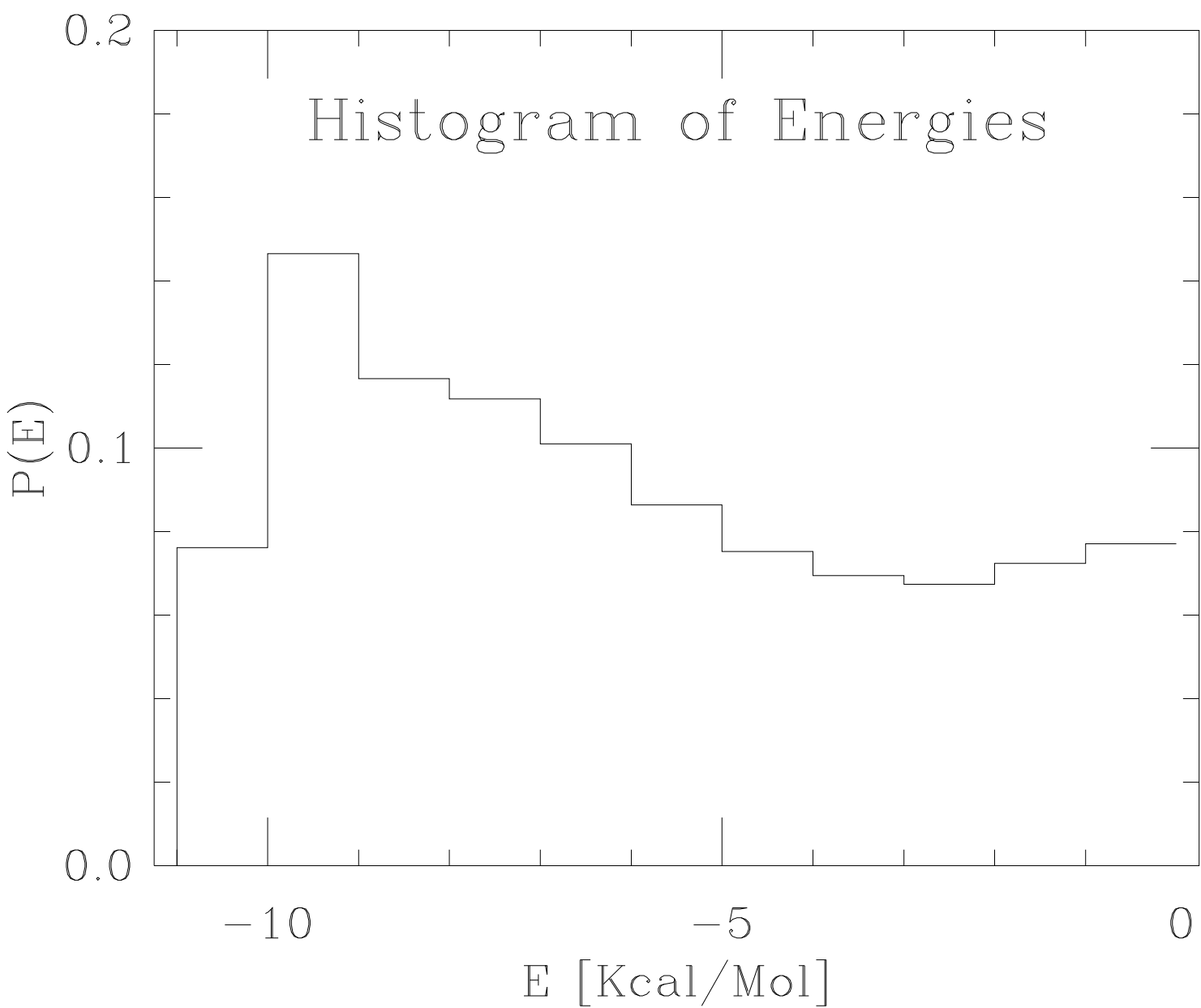
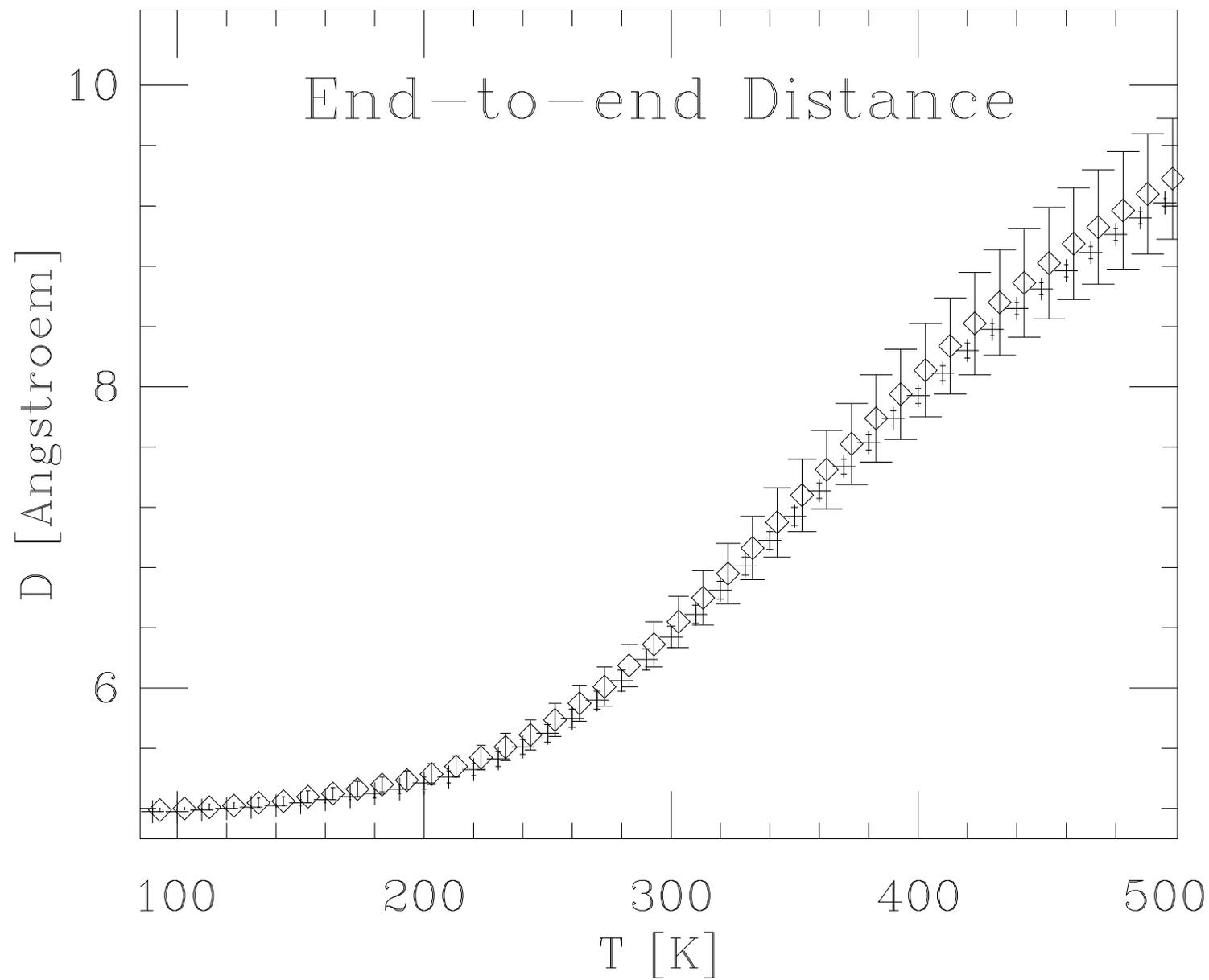
\end{document}